\documentclass[preprintnumbers,amsmath,amssymb,aps,showkeys]{revtex4}

\begin{document}
\title{Mass distribution of highly flattened galaxies and modified Newtonian
dynamics}

\author{
W. F. Kao\thanks{%
gore@mail.nctu.edu.tw} \\
Institute of Physics, Chiao Tung University, HsinChu, Taiwan }

\keywords{dark matter; galaxies: kinematics and dynamics }

\begin{abstract}
Dynamics of spiral galaxies derived from a given surface mass
density has been derived earlier in a classic paper. We try to
transform the singular elliptic function in the integral into a
compact integral with regular elliptic function. Solvable models
are also considered as expansion basis for RC data.  The result
makes corresponding numerical evaluations easier and analytic
analysis possible. It is applied to the study of the dynamics of
Newtonian system and MOND as well. Careful treatment is shown to
be important in dealing with the cut-off of the input data.
\end{abstract}


\maketitle

\section{INTRODUCTION}

 The rotation curve (RC) observations indicates that less than 10
\% of the gravitational mass can be measured from the luminous
part of spiral galaxies. This is the first evidence calling for
the existence of un-known dark matter and dark energy. In the
meantime, an alternative approach, Modified Newtonian dynamics
(MOND) proposed by Milgrom \citep{milgrom}, has been shown to
agree with many rotation curve observations \citep{milgrom,
milgrom1, milgrom2, sanders, sanders1, sanders2}.

Milgrom argues that dark matter is redundant in the approach of
MOND. The missing part was, instead, proposed to be derived from
the conjecture that gravitational field deviates from the
Newtonian $1/r^2$ form when the field strength $g$ is weaker than
a critical value $g_0 \sim 0.9 \times 10^{-8} $ cm s$^{-2}$
\citep{sanders2}.

In MOND the gravitational field is related to the Newtonian
gravitational field $g_N$ by the following relation:
\begin{equation}
\label{1} g \cdot \mu_0 ({g \over g_0}) =g_N
\end{equation}
with a function $\mu_0$ considered as a modified inertial. Milgrom
shows that the model with
\begin{equation} \label{2}
\mu_0 (x) = {x  \over  \sqrt{1+x^2}}
\end{equation}
agrees with RC data of many spiral galaxies \citep{milgrom,
milgrom1, milgrom2, sanders, sanders1, sanders2, kao05}. The
alternative theory could be compatible with the spatial
inhomogeneity of general relativity theory.  Various approaches to
derive the collective effect of MOND has been an active research
interest recently. \citep{mond1, mond2}

{Recently, it was shown that a simpler inertial function of the
following form \citep{fam, fam1}
\begin{equation} \label{03}
\mu_0 (x) = {x  \over  1+x}
\end{equation}
fits better with RC data of Milky Way and NGC3198. Indeed, one can
show that the MOND field strength $g$ is, from Eq. (1),
\begin{equation} \label{04}
g={ \sqrt{{g_N}^2+4g_0g_N} +g_N \over 2} .
\end{equation}
The model shown by Eq. (2) will be denoted as Milgrom model, while
the model (3) will be denoted as FB model. We will study and
compare the results of these two models in the paper. }

{Accumulated evidences show that the theory of MOND is telling us
a very important message. Either the Newtonian force laws do
require modification in the weak field limit or the theory of MOND
may just represent some collective effect of the cosmic dark
matter awaiting for discovery. In both cases, the theory of MOND
deserves more attention in order to reveal the complete underlying
physics hinted by these successful fitting results. Known problem
with momentum conservation has been resolved with alternative
covariant theories\citep{mh, bek, mond1, g05, mond2, romero,
kao06}.

It was also pointed out that the proposed inertial function
$\mu_0$ could be functional of the whole $N$ field and could be
more complicate than the ones shown earlier \citep{bek84, brada,
milgrom3}. Successful fitting with the rotation curves of many
existing spiral galaxies indicates, however, that the most
important physics probably has been revealed by these simple
inertial functions shown in this paper. One probably should take
these models more seriously in order to generate more clues to the
final theory.

If the theory of the MOND is the final theory of gravitation
without any dark matter in a large scale system with some inertial
function $\mu_0(x)$, one should be able to derive the precise mass
distribution form the measured RC provided that the distance $D$
is known. Earlier on, the mass distribution derived from the
rotation curve measurement can only be used to predict how much
dark matter is required in order to secure the Newtonian force
law. In the case of MOND, one should be able to plot a dynamical
profile of the $\Gamma$ function ${\Gamma}(r) \equiv M(r)/L(r)$
with the measured luminosity function $L(r)$ following the theory
of MOND. The $\Gamma$ function should then provide us useful
information about the detailed distribution of stars with
different luminous spectrum obeying the well-known $L \sim
M^\alpha$ relation.

The dynamical profile $\Gamma(r)$ can be treated as important
information regarding the detailed distribution of stars with
different mass-luminosity relations within each spiral
galaxy\citep{jk}. Even one still does not know a reliable way to
derive the inertial function $\mu_0(x)$, there are successful
models shown earlier as useful candidates for the theory of MOND.
It would be interesting to investigate the dependence of these
models with the $M/L$ profiles. Hopefully, information from these
comparisons will provide us clues to the final theory.

Therefore, we will try first to derive the mass density for
Milgrom and Famaey\&Binney (FB) models in details. There are
certain boundary constraints needed to be relaxed for the
asymptotic flat rotation curve boundary condition which is treated
differently in the original derivation of these formulae
\citep{toomre}. It is also important to compare the effects of
exterior contribution between the Milgrom and FB models for a more
precise test of the fitting application.

}

We will first review briefly how to obtain the surface mass
density $\mu(r)$ from a given Newtonian gravitational field $g_N$
with the help of the elliptic function $K(r)$ in section II. The
integral involving the Bessel functions is derived in detailed for
heuristical reasons in this section too.

In addition, a series of integrable model in the case of Newtonian
model, Milgrom model as well as the FB model for MOND will be
presented in section III. These solvable models will be shown to
be good expansion basis for the RC curve data for spiral galaxies.
In practice, this expansion method will help us better understand
the analytic properties of the spiral galaxies.

We will also try to convert the formula shown in section II into a
simpler form making numerical integration more accessible in
section IV. The apparently singular elliptic function $K(r)$ is
also converted to combinations of regular elliptic function $E(r)$
by properly managed integration-by-part.

In section V, one derives the interior mass contribution
$\mu(r<R)$ from the possibly unreliable data $v(r
>R)$ both in the cases of MOND and in the Newtonian dynamics. The
singularity embedded in the useful formula is taken care of with
great caution. Similarly, one tries to derive the formulae related
$g_N$ from a given $\mu(r)$ in section VI. One also presents a
simple model of exterior mass density $\mu(r >R)$ in this section.
The corresponding result in the theory of MOND is also presented
in this paper. Finally, we draw some concluding remarks in section
VII.

\section{Newtonian dynamics of a highly flattened galaxy}
Given the surface mass density $\mu(r)$ of a flattened spiral
galaxy, one can perform the Fourier-Bessel transform to convert
$\mu(r)$ to $\mu(k)$ in $k$-space via the following equations
\citep{arfken}
\begin{eqnarray}
\mu(r)&=& \int_0^\infty k dk \; \mu(k) J_0(kr) , \label{3} \\
\mu(k)&=& \int_0^\infty r dr \;  \mu(r) J_0(kr) \label{4}
\end{eqnarray}
with $J_m(x)$ the Bessel functions. Note that the closure relation
\begin{equation} \label{5}
\int_0^\infty k dk J_m(kx) J_m(kx') = {1 \over x} \delta(x-x')
\end{equation}
can be used to convert Eq. (\ref{4}) to Eq. (\ref{3}) and vice
versa with a similar formula integrating over $dr$. The Green
function of the equation
\begin{equation}
\nabla^2 G (x)= -4 \pi G \delta (r) \delta(z)
\end{equation}
can be read off from the identity
\begin{equation}
{1 \over \sqrt{r^2 +z^2}} =  \int_0^\infty  dk \; \exp[-k |z|]
J_0(kr) .
\end{equation}
Therefore, the Newtonian potential $\phi_N$ can be shown to be
\citep{toomre, kent}
\begin{equation} \phi_N (r,z) = 2 \pi G \int_0^\infty dk \; \mu(k)
J_0(kr)\exp[-k |z|]
\end{equation}
with a given surface mass density $\mu(r)$. It follows that
\begin{equation} \label{9}
g_N(r)= - \partial_r \phi_N (r, z=0) =2 \pi G \int_0^\infty k dk
\; \mu(k) J_1(kr) .
\end{equation}
The Newtonian gravitational field $g_N(r)= v_N^2(r) /r$ can be
readily derived with a given function of rotation velocity
$v_N(r)$. Here one has used the recurrence relation
$J_1(x)=-J_0'(x)$ in deriving Eq. (\ref{9}). Furthermore, from the
closure relation (\ref{5}) of Bessel function, one can show that
the function $g_N(r)$ satisfies the following conversion equation
\begin{equation}
g_N(r) = \int_0^\infty k dk \int_0^\infty r' dr' \; g_N(r')
J_1(kr) J_1(kr') . \label{10}
\end{equation}
Therefore, one has
\begin{equation} \label{11}
\mu(k) = {1 \over 2 \pi G}  \int_0^\infty r dr \;  g_N(r) J_1(kr)
.
\end{equation}


Hence, with a given surface mass density $\mu(r)$ for a flattened
spiral galaxy, one can show that
\begin{equation} \label{121}
\mu(r)={1 \over 2 \pi G} \int_0^\infty k dk \int_0^\infty  dr' \;
r' g_N(r') J_0(kr) J_1(kr') .
\end{equation}

Assuming that $\lim_{r \to 0} rg_N(r) \to 0$ and $\lim_{r \to
\infty} rg_N(r) < \infty $ hold as the boundary conditions, one
can perform an integration-by-part and show that
\begin{equation}
\mu(r)={1 \over 2 \pi G} \int_0^\infty dk \int_0^\infty dr' \;
\partial_{r'} [r'  g_N(r')] J_0(kr) J_0(kr')
\end{equation}
with the help of the asymptotic property of Bessel function:
$J_m(r \to \infty) \to 0$. Here we have used the recurrence
relation $J_1(x)= -J'_0(x)$ in deriving above equation.

Note that the boundary terms can be eliminated under the
prescribed boundary conditions. In fact, the limit $\lim_{r \to
\infty} rg_N(r) < \infty $ is slightly different from the original
boundary conditions given in \citep{toomre}. The difference is
aimed to make the system consistent with the flatten RC
measurement which implies that $\lim_{r \to \infty} rg_N(r) =
v^2_N(r\to \infty) \to $ constant $< \infty$. In summary, one only
needs to modify the asymptotic boundary condition in order to
eliminate the surface term during the process of
integration-by-part. We have relaxed this boundary condition to
accommodate any system with a flat rotation curve.

One can further define
\begin{equation} \label{14}
H(r, r')=\int_0^\infty dk J_0(kr) J_0(kr')
\end{equation}
and write the function $\mu(r)$ as
\begin{equation}
\mu(r)= {1 \over 2 \pi G} \int_0^\infty dr' \; \partial_{r'} [r'
g_N(r')] H(r, r').
\end{equation}
Note that the function $H(r, r')$ can be shown to be proportional
to the elliptic function $K(x)$:
\begin{equation} \label{161}
H(r, r')=  {2
\over \pi r_>}K({r_< \over r_>})
\end{equation}
with $r_>$ ($r_<$) the larger (smaller) of $r$ and $r'$.

The proof is quite straightforward. Since some properties of the
Bessel functions are very important in the dynamics of the spiral
disk, as well as many disk-like system, we will show briefly the
proof for heuristical reason. One of the purpose of this
derivation is to clarify that there are different definitions for
the elliptic functions written in different textbooks. Confusion
may arise applying the formulae in a wrong way.

Note first that the Bessel function $J_0(x)$ has an integral
representation \citep{arfken, bessel}
\begin{equation} \label{171}
J_0(x)={1 \over  \pi} \int_0^\pi \cos [x \sin \theta ]d\theta .
\end{equation}
Given the delta function represented by the plane wave expansion:
\begin{equation}
\delta(x)=  {1 \over 2 \pi} \int_{-\infty}^\infty dk \exp [ikx] ,
\end{equation}
One can show that, with the integral representation of $J_0$,
\begin{equation} \label{19}
 \int_0^\infty dx \cos [kx] \;  J_0(x) = {1 \over \sqrt{1-k^2}}
 \end{equation}
for $k <1$. On the contrary, above integral vanishes for $k >1$.
Therefore, one can apply above equation to show that
\begin{equation} \label{151}
\int_0^\infty dk J_0(kx) J_0(k) = {2 \over  \pi} \int_0^{\pi /2}
{d \theta \over \sqrt{1-x^2 \sin ^2 \theta}} = {2 \over \pi} K(x)
\end{equation}
with the help of the Eq. (\ref{171}) again. The last equality in
above equation follows exactly from the definition of the elliptic
function $K$.

There is an important remark here. Note that the elliptic
functions $E(k)$ and $K(k)$ used in this paper, and Ref.
\citep{toomre, arfken} as well, are defined as
\begin{eqnarray}\label{20}
K(x) &\equiv & \int_0^{\pi/2} (1-x^2 \sin ^2 \theta)^{-1/2} d
\theta ,\\ \label{21} E(x) &\equiv & \int_0^{\pi/2} (1-x^2 \sin ^2
\theta)^{1/2} d \theta
\end{eqnarray}
which is different from certain textbooks. Some textbooks, and
similarly some computer programs, prefer to define above integrals
as $K(x^2)$ and $E(x^2)$ instead. In fact, one can check the
differential equations satisfied by $E(x)$ and $K(x)$ that will be
shown explicitly shortly in next section. Before adopting the
equations presented in the texts, one should check the definition
of these elliptic functions carefully for consistency.

After a proper redefinition of $k \to k'r'$ and write $x=r/r' <1$,
one can readily prove that the assertion (\ref{14}) is correct.

Note that the series expansion of the elliptic function $K(x)$ is
\begin{equation} \label{A}
K(x) = {\pi \over 2} \sum_{n=0}^\infty C_n^2x^{2n}
\end{equation}
with $C_0=1$ and $C_n = (2n-1)!!/2n!!$ for all $n \ge 1$.
Therefore, one can show that it agrees with the series expansion
of the hypergeometric function $F({1 \over 2}, {1 \over 2}, 1;
{r_<^2 \over r_>^2})$, up to a factor $\pi /2$. Indeed, one has:
\begin{equation} \label{B}
F(a,b,c; {x^2}) = \sum_{n=0}^\infty  {(a)_n (b)_n \over (c)_n n!}
x^{2n},
\end{equation}
with $(a)_n \equiv \Gamma(n+a)/\Gamma(a)$. It is straightforward
to show that the expansion coefficients of Eq.s (\ref{A}) and
(\ref{B}) agree term by term. Therefore, the derivation shown
above agrees with the result shown in \citep{bessel, toomre}:
\begin{equation}
H(r,r')={1 \over r_>} F({1 \over 2}, {1 \over 2}, 1; {r_<^2 \over
r_>^2}) = {2 \over \pi r_>}K({r_< \over r_>}).
\end{equation}
with $r_>$ ($r_<$) the larger (smaller) of $r$ and $r'$.

As a result, one has
\begin{equation} \label{17}
\mu(r)={1 \over 2 \pi G} \int_0^\infty \partial_{r'} [v_N^2(r')]
\; H(r, r')dr'
\end{equation}
given the identification $g_N = v_N^2/r$.

\section{Some Simple integrable models}

One can show that $\mu(r)$ is integrable with a Newtonian velocity
described by
\begin{equation} \label{vN2}
v_N^2(r) = {C_0^2 a \over \sqrt{ r^2 + a^2}},
\end{equation}
with $C_0$ and $a$ some constants of parametrization.  Note that
the velocity function $v_N(r)$ vanishes at spatial infinity while
$v_N(0) \to C_0$ with a non-vanishing value. One notes that the
observed velocity at $r=0$ is expected to be zero from symmetric
considerations. We will come back to this point later for a
resolution. Following Eq. (\ref{121}), one can write
\begin{equation} \label{12}
\mu(r)={1 \over 2 \pi G} \int_0^\infty k dk \Lambda(k) J_0(kr)
\end{equation}
with $\Lambda(k)$ defined as
\begin{equation} \label{lamdak}
\Lambda(k) \equiv  \int_0^\infty {C_0^2 a \over \sqrt{r^2 +
a^2}}J_1(kr)dr.
\end{equation}
To evaluate $\Lambda(k)$, one will need the following formula:
\begin{equation} \label{31}
\int_0^\infty dk \exp [-kx] J_0(k)={1 \over  \sqrt{1+ x^2}}
\end{equation}
which follows from Eq. (\ref{19}) by replacing $x \to ix$ with the
help of an analytic continuation. In addition, writing $ \exp
[-kx]\, dk$ as $-d (\exp [-kx])/x$ and performing an
integration-by-part, one can derive
\begin{equation} \label{311}
\int_0^\infty dk \exp [-kx] J_1(k)=1- {x \over  \sqrt{1+ x^2}}.
\end{equation}
Here we have used the identity $J_0'=-J_1$ and the fact that
$J_0(0)=1$. With the help of above equation, one can show that
\begin{equation}
\int_0^\infty dk (1-\exp [-2k]) J_1(kx)= {2 \over x \sqrt{x^2+4}
}.
\end{equation}
Multiplying both sides of above equation with $J_1(k'x)xdx$, one
can integrate above equation and obtain
\begin{equation}
\int_0^\infty dx J_1(kx) {2 \over \sqrt{x^2+4}} = {1- \exp[-2k]
\over k }
\end{equation}
with the help of the closure relation (\ref{5}). After a
redefinition of parameters $k \to ka/2$ and $x \to 2r/a$, one can
show that
\begin{equation}
\int_0^\infty dr J_1(kr) {a \over \sqrt{r^2+a^2}} = {1- \exp[-ak]
\over k }
\end{equation}
and hence
\begin{equation}
\Lambda(k)=  {C_0^2 \over k} \left ( 1-\exp [-ak] \right).
 \end{equation}
In additon, equation (\ref{31}) can also be written as
\begin{equation} \label{31_1}
\int_0^\infty dk \exp [-kx] J_0(ka)={1 \over  \sqrt{ x^2+a^2}}
\end{equation}
after proper reparametrization. Therefore, one has
\begin{eqnarray}
\mu(r)&=&{C_0^2 \over 2 \pi G} \int_0^\infty dk (1-\exp [-ak])
 J_0(kr) \nonumber \\
&=& {C_0^2 \over 2 \pi G}  \left [ {1 \over r} - {1 \over \sqrt{
r^2 + a^2}} \right ]. \label{38_05}
\end{eqnarray}
 Hence this model is integrable as promised. Also, as
mentioned earlier in this section, this model with a small $C_0$
and properly adjusted $a$ can describe the velocity profile pretty
nice in the case of MOND.

One can eliminate the non-vanishing constant by two different
methods. The first method is simply subtracting two integrable
$v_N^2(r, C_0, a_i)$, namely, define the new Newtonian velocity as
\begin{equation}\label{vN22}
v_{N0}^2(r) = C_0^2\left( {a_1 \over \sqrt{ r^2 + a_1^2}} -{a_2
\over \sqrt{ r^2 + a_2^2}} \right ),
\end{equation}
with $C_0$ and $a_1 > a_2$ some constants of parameterizations.
This new velocity function is also integrable due to the linear
dependence of the function $v_N^2(r)$ in Eq. (\ref{17}). In
addition, it vanishes at $r=0$ and approaches $0$ at spatial
infinity $r \to \infty$. Therefore, one has
\begin{eqnarray}
\mu_0(r)&=& {C_0^2 \over 2 \pi G}  \left [ {1 \over \sqrt{ r^2 +
a_2^2}} - {1 \over \sqrt{ r^2 + a_1^2}} \right ] \label{38_051}
\end{eqnarray}
directly from Eq. (\ref{38_05}). In addition, one can also show
that \citep{toomre} higher derivative models defined by
\begin{equation}
v_{Nn}^2(r) =-C_n^2 (-{ \partial \over \partial a^2})^n {a \over
\sqrt{ r^2 + a^2}}= \sum_{k=1}^n  {C^n_k (2k-1)!  (2n-2k)!
 \over 2^{2n-1} (k-1)!(n-k)! (2k-1)
}a^{1-2k}(r^2 + a^2)^{-n+k-1/2}-{(2n)! \over 2^{2n} n! }a (r^2 +
a^2)^{-n+1/2}
\end{equation}
are also integrable and give the mass density as
\begin{eqnarray}
\mu_n(r)&=& -{C_n^2 \over 2 \pi G} (-{ \partial \over \partial
a^2})^n \left [ {1 \over r} - {1 \over \sqrt{ r^2 + a^2}} \right
]={C_n^2 \over 2 \pi G} (-{ \partial \over \partial a^2})^n \left
[  {1 \over \sqrt{ r^2 + a^2}} \right ] = {C_n^2 \over 2 \pi G}
{(2n)! \over 2^{2n} n! } (r^2 + a^2)^{-n-1/2}.
\end{eqnarray}
Mote that both $v_{Nn}^2$ and $\mu_n(r)$ are in fact functions of
$(r^2 + a^2)^{-n-1/2}$ with appropriate combinations.  For
example, given the Newtonian velocity
\begin{equation} v_{N1}^2(r)
={C_1^2 r^2 \over a ( r^2 + a^2)^{3/2}}= C_1^2 \left [ { 1 \over a
( r^2 + a^2)^{1/2}} - { a \over ( r^2 + a^2)^{3/2}} \right ],
\label{42_05}
\end{equation}
the corresponding mass density will be given by
\begin{equation} \label{tmmu1}
\mu_1(r)=    {C_1^2 \over 2 \pi G( r^2 + a^2)^{3/2}}.
\end{equation}
For convenience, we have absorbed a common factor $1/2$ into
$C_1^2$. Note that $v_{N1}^2(r \to \infty) \to 0$ and $v_1(r=0)=0$
in this model.

\subsection{Newtonian Model}
Consider the case of Newtonian model that observed velocity $v$
and Newtonian velocity $v_N$ are identical. This model is known to
require the presence of dark matters\cite{gentile}. Since the
velocity has to vanish at $r=0$ and goes to a constant at spatial
infinity, one will show that an additional constant term added to
the $v_N^2$ will provide both resolution at the same time. Indeed,
another way to eliminate the non-vanishing velocity at $r=0$ is to
introduce a constant velocity by noting that
\begin{equation} \label{12}
\mu(r)={1 \over 2 \pi G} \int_0^\infty k dk \Lambda(k)
J_0(kr)={C_0^2 \over 2 \pi G r}
\end{equation}
with $\Lambda(k)$ defined as
\begin{equation} \label{lamdak}
\Lambda(k) \equiv C_0^2 \int_0^\infty J_1(kr)dr={C_0^2 \over k}.
\end{equation}
Here we have used the identity
\begin{equation} \label{Jnkr}
\int_0^\infty dk J_n(kr)={1 \over  r}
\end{equation}
which follows directly from Eq. (\ref{31}-\ref{311}).

Therefore, one can show that the Newtonian velocity given by the
form
\begin{equation} \label{vN2N}
v_0^2(r)= v_{N0}^2(r) =C_0^2\left[ 1- { a \over \sqrt{ r^2 + a^2}}
\right ]
\end{equation}
is induced by the mass density of the following form
\begin{eqnarray}
\mu_0(r)&=& {C_0^2 \over 2 \pi G}  {1 \over \sqrt{ r^2 + a^2}}.
\label{38_05N}
\end{eqnarray}
Here $C_0$ and $a$ are some constants of parametrization. Note
that the Newtonian velocity (\ref{vN2N}) vanishes at $r=0$ and
$V_N \to C_0$ at spatial infinity as promised. Therefore, this
velocity profile goes along with the observed RC of the spirals.
Hence this model can be used to simulate the dynamics of spiral
galaxies which requires the presence of dark matters. Hence we
will call this model with  $v=v_N$is the Newtonian model that
normally requires the existence of dark matters.

This zero mode is however the only known integrable modes for
velocity profiles. Successive differentiating $v_N^2$ in this
model simply turns off the constant term $C_0^2$ in Eq.
(\ref{vN2N}). Therefore, higher derivative modes derived from
further differentiation of the Newtonian potential $v_N^2$ with
respect to $-a^2$,
\begin{equation}
 v_{n}^2 \equiv {  C_n^2 \over  C_0^2 }(-\partial_{a^2})^n v_0^2=
 C_n^2  (-\partial_{a^2})^n \left[  { a \over \sqrt{ r^2 + a^2}} \right
 ]
\end{equation}
will simply take away the leading constant term from the higher
order modes. Therefore, velocity profiles $v_n(r)$ will be quite
different from $v_0$ in this case. Note again that the velocity
$v_n$ is induced by the corresponding mass density
\begin{eqnarray}
\mu_n(r) = {C_n^2 \over 2 \pi G} {(2n-1)!! \over 2^n } (r^2 +
a^2)^{-n-1/2}.
\end{eqnarray}
One of the advantage of these well-behaved smooth velocity
functions is that they can be used as expansion basis for
simulation of velocity profiles. Thanks to the linear dependence
of the $v_N^2(r)$ in the mass density function $\mu(r)$, one can
freely combine any integrable modes of velocity to obtain all
possible combinations of integrable models. For example, the model
with
\begin{equation}
v^2(r) = v_N^2(r)= \sum_{i,j} v_{0}^2(r,C_{0i}, a_j) \equiv
\sum_{i,j} C_{0i}^2\left[ 1- { a_j \over \sqrt{ r^2 + a_j^2}}
\right ]
\end{equation}
is integrable and can be shown to be derived by the mass density
\begin{equation}
\mu(r)= \sum_i \mu_0(r, C_{0i}, a_j) \equiv \sum_i {C_{0i}^2 \over
2 \pi G} {1 \over \sqrt{ r^2 + a_j^2}}
\end{equation}
with $C_{ni}$ and $a_j$ all constants of parametrization. Here
summation over $i,j$ is understood to be summed over all possible
modes with different spectrum described by $C_{0i}$ and $a_j$.
This velocity vanishes at $r=0$ and goes to
\begin{equation}
 v^2(r)\to \sum_{i,j} C_{0i}^2
\end{equation}
at spatial infinity. One can also add higher derivative terms
$v_n^2(r)$ to the velocity profile $v^2(r)$. Since higher
derivative velocity $v_n$ vanishes both at $r=0$ and spatial
infinity, these additional terms will not affect the asymptotic
behavior of $v^2$ at $r=0$. Therefore, one needs to keep at least
one zeroth order term in order for $v$ to be compatible with the
RC data.

To be more specific, one can consider the model
\begin{equation}
v^2(r) = v_N^2(r)= \sum_{i,j,n} v_{n}^2(r,C_{ni}, a_j)\equiv
\sum_{i,j} C_{0i}^2\left[ 1- { a_j \over \sqrt{ r^2 + a_j^2}}
\right ]+ \sum_{k,l,n} C_{nk}^2 (-\partial_{a_l^2})^n \left[ { a_l
\over \sqrt{ r^2 + a_l^2}} \right ]
\end{equation}
which be derived by the mass density
\begin{equation}
\mu(r)= \sum_i \mu_n(r, C_{ni}, a_j) \equiv \sum_{i,j} {C_{0i}^2
\over 2 \pi G} {1 \over \sqrt{ r^2 + a_j^2}} + \sum_{n,k,l}
{C_{nk}^2 \over 2 \pi G} {(2n-1)!! \over 2^n } (r^2 +
a_l^2)^{-n-1/2}
\end{equation}
with $C_{0i}$ and $a_j$ are all constants of parameterization.
Here the summation over $n$ is to be summed over all $n \ge 1$.
The velocity of these models will vanish at $r=0$ and goes to
\begin{equation}
 v^2(r)\to \sum_{i,j} C_{0i}^2
\end{equation}
at spatial infinity. Therefore, these models turn out to be good
expansion basis of any RC data for Newtonian dynamics.

In practice, one may fit the $v^2$ in expansion of these modes in
order to analyze the RC in basis of of these basis modes. This
helps analytical understanding of the spiral galaxies more
transparently. The properties of each modes is easy understand
because they are integrable. The corresponding coefficients
$C_{ni}$ and $a_j$ will determine the contributions of each modes
to any galaxies. One will be able to construct tables for spiral
galaxies with the corresponding coefficients of each modes.
Hopefully, this expansion method originally developed in
\cite{toomre} will provide us a new way to look at the major
dynamics of the spiral galaxies.

\subsection{Milgrom model}

For the case of Milgrom model (2), the Newtonian velocity $v_N$
and the observed velocity $v$ are related by the following
equation
\begin{equation} \label{Mv}
v^2(r) = \left [{{V_N^4(r) + \sqrt{V_N^8(r) +4 V_N^4(r)g_0^2r^2}
\over 2 } }\right ]^{1/2}.
\end{equation}
Therefore, one can show that a galaxy with a rotation curve given
by
\begin{equation}
v_0^4(r)= {C_0^4a^2 \over 2( r^2 + a^2)} \left \{ 1 +  \sqrt{ 1+ {
4g_0^2 r^2 ( r^2 + a^2)  \over C_0^4a^2 } } \right \}
\end{equation}
is the corresponding velocity induced by the Newtonian velocity
\begin{equation} \label{vN2}
v_{N0}^2(r) = {C_0^2 a \over \sqrt{ r^2 + a^2}}.
\end{equation}
Here $C_0$ and $a$ are some constants of parametrization.
Therefore, the corresponding mass density is given by Eq.
(\ref{38_05})
$$ \mu_0(r)= {C_0^2 \over 2 \pi G}  \left [ {1 \over r} - {1 \over \sqrt{
r^2 + a^2}} \right ]. $$
Note that $v_0^4$ approaches a constant
$g_0 C_0^2a$ at spatial infinity. This asymptotically flat pattern
of the velocity is compatible with many observations of the
spirals. This Newtonian velocity does not, however, vanish at the
origin. Indeed, one can show that $v(0) \to C_0 \ne 0$ Therefore
this would not be a good expansion basis for the physical spirals.
The non-vanishing behavior of $v$ at $r=0$ can be secured by
considering the refined model (\ref{vN22}):
$$ v_{N0}^2(r) = C_0^2\left( {a_1 \over \sqrt{ r^2 + a_1^2}} -{a_2 \over
\sqrt{ r^2 + a_2^2}} \right )$$ with $C_0$, $a_1>a_2$ some
constants of parameterizations. The corresponding velocity
function $v$ can be shown to be
\begin{equation}
v_0^4(r)= {C_0^4 \over 2} \left [ { a_1 \over \sqrt{ r^2 +
a_1^2}}- { a_2 \over \sqrt{ r^2 + a_2^2}} \right ]^2 \left \{ 1 +
\left [{ 1+ { 4g_0^2 r^2 ( r^2 + a_1^2) ( r^2 + a_2^2) \over C_0^4
\left (a_1\sqrt{ r^2 + a_2^2}- a_2\sqrt{ r^2 + a_1^2} \right )^2 }
} \right ]^{1/2} \right \}
\end{equation}
This new velocity function $v$ is hence induced by the Newtonian
velocity (\ref{vN22}). In addition, Note that $v^4$ approaches a
constant $g_0 C_0^2(a_1-a_2)$ at spatial infinity and vanishes at
$r=0$. This agrees with the main feature of the observed
asymptotically flat rotation curve of many spirals. As a result,
the corresponding mass density of this model is given by Eq.
(\ref{38_051}):
$$ \mu_0(r)= {C_0^2 \over 2 \pi G}  \left [ {1 \over \sqrt{ r^2 +
a_2^2}} - {1 \over \sqrt{ r^2 + a_1^2}} \right ].$$ In addition, a
model with a velocity, in the case of Milgrom model, of the form
\begin{equation} \label{tmvn1}
v_1^4 = { C_1^4 r^4 + \sqrt{C_1^8 r^8 + 4 C_1^4 r^6 g_0^2
a^2(r^2+a^2)^3 } \over  2 a^2(r^2+a^2)^3}
\end{equation}
can be shown to be induced by the Newtonian velocity of the
following form given by Eq. (\ref{42_05}):
\begin{equation} v_{N1}^2(r)
={C_1^2 r^2 \over a ( r^2 + a^2)^{3/2}} . \nonumber
\end{equation}
Therefore, this model is derived by the mass density
(\ref{tmmu1}):
\begin{equation} \nonumber
\mu_1(r)=    {C_1^2 \over 2 \pi G( r^2 + a^2)^{3/2}}.
\end{equation}
Note that $v_1^2(r \to \infty) \to C_1 \sqrt{g_0/a}$ and
$v_1(r=0)=0$ in this Milgrom model. In addition, the corresponding
Newtonian model also has the same properties: $v_{N1}(r)$ vanishes
both at $r=0$ and $r \to \infty$. Therefore, this model also
appears to be a realistic model in agreement with the
asymptotically flat rotation curve being observed.

Note again that further differentiation of the Newtonian velocity
$v_N^2$ with respect to $-a^2$ will derive integrable higher
derivative models:
\begin{equation}
 v_{Nn}^2 \equiv
 C_n^2  (-\partial_{a^2})^n \left[  { a \over \sqrt{ r^2 + a^2}} \right
 ].
\end{equation}
Therefore, this velocity will be derived by the mass density
\begin{eqnarray}
\mu_n(r) = {C_n^2 \over 2 \pi G} {(2n-1)!! \over 2^n } (r^2 +
a^2)^{-n-1/2}.
\end{eqnarray}

One of the advantage of these well-behaved smooth velocity
functions is that they can be used as expansion basis for
simulation of velocity profiles. Thanks to the linear dependence
of the $v_N^2(r)$ in the mass density function $\mu(r)$, one can
freely combine any integrable modes of velocity to obtain all
possible combinations of integrable models. For example, the model
with
\begin{equation}
 v_N^2(r)= \sum_{i,j} v_{N0}^2(r,C_{0i}, a_j, b_j) \equiv
\sum_{i,j} C_{0i}^2\left[{ a_j \over \sqrt{ r^2 + a_j^2}}- { b_j
\over \sqrt{ r^2 + b_j^2}} \right ]
\end{equation}
is integrable and can be shown to be derived by the mass density
\begin{equation}
\mu(r)= \sum_i \mu_0(r, C_{0i}, a_j) \equiv \sum_i {C_{0i}^2 \over
2 \pi G} \left [ {1 \over \sqrt{ r^2 + b_j^2}} - {1 \over \sqrt{
r^2 + a_j^2}} \right ]
\end{equation}
with $C_{ni}$ and $a_j$ all constants of parametrization. The
velocity $v_N^2$ also vanishes at $r=0$ and goes to
\begin{equation}
 v_N^2(r)\to \sum_{i,j} C_{0i}^2 {a_j-b_j \over r}
\end{equation}
at spatial infinity. This will in turn make the corresponding
observed Milgrom velocity $v^2$ approaches the asymptotic velocity
$v^2_\infty \to [\sum_{i,j}C_{0i}^2g_0 (a_j-b_j)]^{1/2}$. One can
also adds higher derivative terms $v_{Nn}^2(r)$ to the velocity
profile $v_N^2(r)$. Since higher derivative velocity $v_n$ goes to
zero faster than the zero-th derivative term at spatial infinity,
these adding will not affect the asymptotic behavior of $v^2$ at
spatial infinity. Therefore, leading order terms will determine
the asymptotic value of $v$. To be more specific, one can consider
the model
\begin{equation}
v_N^2(r)= \sum_{i,j,n} v_{n}^2(r,C_{ni}, a_j)\equiv \sum_{i,j}
C_{0i}^2\left[ { a_j \over \sqrt{ r^2 + a_j^2}}- { b_j \over
\sqrt{ r^2 + b_j^2}} \right ]+ \sum_{k,l,n} C_{nk}^2
(-\partial_{a_l^2})^n \left[ { a_l \over \sqrt{ r^2 + a_l^2}}
\right ]
\end{equation}
which be derived by the mass density
\begin{equation}
\mu(r)= \sum_i \mu_n(r, C_{ni}, a_j) \equiv \sum_{i,j} {C_{0i}^2
\over 2 \pi G} \left [ {1 \over \sqrt{ r^2 + b_j^2}} - {1 \over
\sqrt{ r^2 + a_j^2}} \right ] + \sum_{n,k,l} {C_{nk}^2 \over 2 \pi
G} {(2n-1)!! \over 2^n } (r^2 + a_l^2)^{-n-1/2}
\end{equation}
with $C_{0i}$ and $a_j$ all constants of parametrization. Note
again that the velocity $v^2$ of these models also vanishes at
$r=0$ and goes to
\begin{equation}
 v_N^2(r)\to \sum_{i,j} C_{0i}^2 {a_j-b_j \over r}
\end{equation}
corresponding to
\begin{equation}
 v^2(r)\to \left [g_0 \sum_{i,j} C_{0i}^2 {(a_j-b_j)} \right ]^{1/2}
\end{equation}
 at spatial infinity. Therefore, these models
turn out to be good expansion basis for $v_N^2(r)$ of any RC data
for Milgrom models.

In practice, one may convert the RC data from $v$ to $v_N$
following Eq. (\ref{Mv}) and then fit the resulting $v_N^2$ in
expansion of these modes in order to analyze the RC in basis of of
these basis modes. This helps analytical understanding of the
spiral galaxies more transparently. The properties of each modes
is easy understand because they are integrable. The corresponding
coefficients $C_{ni}$ and $a_j$ will determine the contributions
of each modes to any galaxies. One will be able to construct
tables for spiral galaxies with the corresponding coefficients of
each modes. Hopefully, this expansion method originally developed
in \cite{toomre} will provide us a new way to look at the major
dynamics of the spiral galaxies.

\subsection{Famaey and Binney model}
For the case of FB model (\ref{03}), the Newtonian velocity $v_N$
and the observed velocity $v$ are related by the following
equation
\begin{equation} \label{FB04v}
v^2={ \sqrt{{v_N}^4+4g_0rv_N^2} +v_N^2 \over 2} .
\end{equation}
Therefore, one can show that a galaxy with a rotation curve given
by
\begin{equation} \label{44_05}
v_0^2(r)= {C_0^2a \over 2\sqrt{r^2 + a^2}} \left \{ 1 +  \left [ {
1+ { 4g_0 r \sqrt{ r^2 + a^2}  \over C_0^2a } } \right ]^{1/2}
\right \}
\end{equation}
is the corresponding velocity induced by the Newtonian velocity
\begin{equation} \label{vN2}
v_{N0}^2(r) = {C_0^2 a \over \sqrt{ r^2 + a^2}}.
\end{equation}
Here $C_0$ and $a$ some constants of parametrization. Therefore,
the corresponding mass density is given by Eq. (\ref{38_05})
$$ \mu_0(r)= {C_0^2 \over 2 \pi G}  \left [ {1 \over r} - {1 \over \sqrt{
r^2 + a^2}} \right ]. $$ Note that $v_0^4$ approaches a constant
$g_0 C_0^2a$ at spatial infinity. This asymptotic flat pattern of
the velocity is compatible with many observations of the spirals.
This Newtonian velocity does not, however, vanish at the origin.
Indeed, one can show that $v_0(0) \to C_0 \ne 0$. Therefore this
would not be a good expansion basis for most physical spirals. The
non-vanishing behavior of $v$ at $r=0$ can be secured by
considering the refined model (\ref{vN22}):
$$ v_{N0}^2(r) = C_0^2\left( {a_1 \over \sqrt{ r^2 + a_1^2}} -{a_2 \over
\sqrt{ r^2 + a_2^2}} \right )$$ with $C_0$, $a_1>a_2$ some
constants of parametrization. The corresponding velocity function
$v$ can be shown to be
\begin{equation}
v_0^2(r)= {C_0^2 \over 2} \left [ { a_1 \over \sqrt{ r^2 +
a_1^2}}- { a_2 \over \sqrt{ r^2 + a_2^2}} \right ] \left \{ 1 +
\left [{ 1+ { 4g_0 r ( r^2 + a_1^2)^{1/2} ( r^2 + a_2^2)^{1/2}
\over C_0^2 \left (a_1\sqrt{ r^2 + a_2^2}- a_2\sqrt{ r^2 + a_1^2}
\right ) } } \right ]^{1/2} \right \}
\end{equation}
This new velocity function $v_0$ is hence induced by the Newtonian
velocity (\ref{vN22}). In addition, Note that $v_0^4$ approaches a
constant $g_0 C_0^2(a_1-a_2)$ at spatial infinity and vanishes at
$r=0$. This fits the main feature of the asymptotically flat
rotation curve of the spirals. As a result, the corresponding mass
density of this model is given by Eq. (\ref{38_051}):
$$ \mu_0(r)= {C_0^2 \over 2 \pi G}  \left [ {1 \over \sqrt{ r^2 +
a_2^2}} - {1 \over \sqrt{ r^2 + a_1^2}} \right ].$$ In addition, a
model with a velocity, in the case of FB model, of the form
\begin{equation}
v_1^2 = { C_1^2 r^2 \over  2 a(r^2+a^2)^{3/2}} \left [ 1 + \left
[{1 + { 4 g_0 a(r^2+a^2)^{3/2} \over C_1^2 r} } \right ]^{1/2}
\right ]
\end{equation}
can be shown to be induced by the Newtonian velocity of the
following form given by Eq. (\ref{42_05}):
\begin{equation} v_{N1}^2(r)
={C_1^2 r^2 \over a ( r^2 + a^2)^{3/2}} . \nonumber
\end{equation}
Therefore, this model is derived by the mass density
(\ref{tmmu1}):
\begin{equation} \nonumber
\mu_1(r)=    {C_1^2 \over 2 \pi G( r^2 + a^2)^{3/2}}.
\end{equation}
Note that $v_1^2(r \to \infty) \to C_1 \sqrt{g_0/a}$ and
$v_1(r=0)=0$ in this FB model. The corresponding Newtonian model
also has the same limit: $v_{N1}(r)$ goes to $0$ in both $r=0$ and
$r \to \infty$ limits. Therefore, this model appears to be a more
realistic model compatible with the flat rotation curve being
observed.

Note again that further differentiation of the Newtonian velocity
$v_N^2$ with respect to $-a^2$ will derive integrable higher
derivative models:
\begin{equation}
 v_{Nn}^2 \equiv
 C_n^2  (-\partial_{a^2})^n \left[  { a \over \sqrt{ r^2 + a^2}} \right
 ].
\end{equation}
Therefore, this velocity can be shown to be derived from the mass
density
\begin{eqnarray}
\mu_n(r) = {C_n^2 \over 2 \pi G} {(2n-1)!! \over 2^n } (r^2 +
a^2)^{-n-1/2}.
\end{eqnarray}

One of the advantage of these well-behaved smooth velocity
functions is that they can be used as expansion basis for
simulation of velocity profiles. Thanks to the linear dependence
of the $v_N^2(r)$ in the mass density function $\mu(r)$, one can
freely combine any integrable modes of velocity to obtain all
possible combinations of integrable models. For example, the model
with
\begin{equation}
 v_N^2(r)= \sum_{i,j} v_{N0}^2(r,C_{0i}, a_j, b_j) \equiv
\sum_{i,j} C_{0i}^2\left[{ a_j \over \sqrt{ r^2 + a_j^2}}- { b_j
\over \sqrt{ r^2 + b_j^2}} \right ]
\end{equation}
is integrable and can be shown to be derived by the mass density
\begin{equation}
\mu(r)= \sum_i \mu_0(r, C_{0i}, a_j) \equiv \sum_i {C_{0i}^2 \over
2 \pi G} \left [ {1 \over \sqrt{ r^2 + b_j^2}} - {1 \over \sqrt{
r^2 + a_j^2}} \right ]
\end{equation}
with $C_{ni}$ and $a_j>b_j$ are all constants of
parameterizations. The velocity will vanish at $r=0$ and goes to
\begin{equation}
 v_N^2(r)\to \sum_{i,j} C_{0i}^2 {a_j-b_j \over r}
\end{equation}
at spatial infinity. This will in turn make the corresponding
observed FB velocity $v$ approach the asymptotic velocity
$v^2_\infty \to [\sum_{i,j}C_{0i}^2g_0 (a_j-b_j)]^{1/2}$. One can
also adds higher derivative terms $v_{Nn}^2(r)$ to the velocity
profile $v_N^2(r)$. Since higher derivative velocity $v_n$ goes to
zero faster than the zero-th derivative term at spatial infinity,
these adding will not affect the asymptotic behavior of $v^2$ at
spatial infinity. Therefore, the leading order terms will
determine the asymptotic behavior of the RC.

To be more specific, one can consider the model
\begin{equation}
v_N^2(r)= \sum_{i,j,n} v_{n}^2(r,C_{ni}, a_j)\equiv \sum_{i,j}
C_{0i}^2\left[ { a_j \over \sqrt{ r^2 + a_j^2}}- { b_j \over
\sqrt{ r^2 + b_j^2}} \right ]+ \sum_{n,k,l} C_{nk}^2
(-\partial_{a_l^2})^n \left[ { a_l \over \sqrt{ r^2 + a_l^2}}
\right ]
\end{equation}
which be derived by the mass density
\begin{equation}
\mu(r)= \sum_i \mu_n(r, C_{ni}, a_j) \equiv \sum_{i,j} {C_{0i}^2
\over 2 \pi G} \left [ {1 \over \sqrt{ r^2 + b_j^2}} - {1 \over
\sqrt{ r^2 + a_j^2}} \right ] + \sum_{n,k,l} {C_{nk}^2 \over 2 \pi
G} {(2n-1)!! \over 2^n } (r^2 + a_l^2)^{-n-1/2}
\end{equation}
with $C_{0i}$ and $a_j$ all constants of parameterization. The
velocity of these models vanishes at $r=0$ and goes to
\begin{equation}
 v_N^2(r)\to \sum_{i,j} C_{0i}^2 {a_j-b_j \over r},
\end{equation}
corresponding to
\begin{equation}
 v^2(r)\to \left [g_0 \sum_{i,j} C_{0i}^2 {(a_j-b_j)} \right ]^{1/2}
\end{equation}
 at spatial infinity. Therefore, these models turn
out to be good expansion basis for $v_N^2$ of any RC data for FB
models.

In practice, one may convert the RC data from $v$ to $v_N$
following Eq. (\ref{FB04v}) and then fit the resulting $v_N^2$ in
expansion of these modes in order to analyze the RC in basis of of
these basis modes. This helps analytical understanding of the
spiral galaxies more transparently. The properties of each modes
is easy understand because they are integrable. The corresponding
coefficients $C_{ni}$ and $a_j$ will determine the contributions
of each modes to any galaxies. One will be able to construct
tables for spiral galaxies with the corresponding coefficients of
each modes. Hopefully, this expansion method originally developed
in \cite{toomre} will provide us a new way to look at the major
dynamics of the spiral galaxies.

\section{compact and regular expression}

In order to put the integral in a numerically accessible form, Eq.
(\ref{17}) for $\mu(r)$ can be written as a more compact form with
a compact integral domain $x \in [0,1]$:
\begin{eqnarray}\label{18}
&& \mu(r)={1 \over \pi^2 G r}\\ && \times \left [ \int_0^1
\partial_x [v_N^2(rx)] \; K(x)dx \nonumber
- \int_0^1
\partial_y [v_N^2({r \over y})] \; K(y)y dy \right ].
\end{eqnarray}
Here one has replaced $x=r'/r$ and $y=r/r'$ in above integral. One
of the advantage of this expression is the numerical analysis
involves only a compact integral domain $0 \le x \le 1$ instead of
a open and infinite domain $0 \le r \to \infty$ domain. Even most
integral vanish quickly enough without bothering the large $r$
domain, the compact expression will make both the numerical and
analytical implication more transparent to access.

Note that the function $K(x)$ diverges at $x=1$.  It is, however,
easy to show that $K(x)dx \to 0$ near $x=1$. Usually, one can
manually delete the negligible integration involving the
elliptical function $K(x \to 1)$ to avoid computer break-down due
to the apparent singularity.

It will be, however, easier for us to perform analytic and/or
numerical analysis with an equation that is free of any apparently
singular functions in the integrand. Indeed, this can be done by
transforming the singular elliptic function $K(x)$ to regular
elliptic function $E(x)$. The advantage of this transformation
will be used to evaluate approximate result in the next sections.
Therefore, one will try to convert the apparently singular
function $K(x)$ into a singular free function $E(x)$ by performing
some proper integration-by-part.

One will need a few identities satisfied by the elliptic functions
$E$ and $K$. Indeed, it is straightforward to show that $E(x)$ and
$K(x)$ satisfy the following equations that will be useful in
converting the integrals into more accessible form:
\begin{eqnarray} \label{22}
x (1-x^2) K''(x)+(1-3x^2)K'(x)-x K(x)&=&0 ,\\
x (1-x^2) E''(x)+(1-x^2)E'(x)+x E(x)&=&0, \label{23}
\end{eqnarray}
and also
\begin{eqnarray}
x E'(x)+K(x)&=&E(x),  \label{24} \\
E'(x)+{x \over 1-x^2} E(x)&=& K'(x), \label{25} \\
x K'(x)+K(x)&=& {1 \over 1-x^2} E(x). \label{26}
\end{eqnarray}
Note that Eq. (\ref{24}) can be derived directly from
differentiating the definition of the elliptic integrals Eq.s
(\ref{20}-\ref{21}). In addition, Eq.s (\ref{25}-\ref{26}) can
also be derived with the help of the Eq. (\ref{23}).

When one is given a set of data as a numerical function of
$v_N(rx)$, it is much easier to compute $dv_N(k=rx)/dk$ instead of
$\partial_x v_N(rx)$. Therefore, one will need the following
converting formulae:
$$ \partial_x [v_N^2(rx)]= r [v_N^2(rx)]',$$
$$ \partial_y [v_N^2({r \over y})]= -{r \over y^2} [v_N^2({r \over
y})]'.$$

Therefore, one is able to write the Eq. (\ref{18}) as:
\begin{equation}
\mu(r)={1 \over \pi^2 G } \left [ \int_0^1 dv_N^2(rx) \; K(x)dx +
\int_0^1 dv_N^2({r \over y}) \; K(y) y^{-1} dy \right ]
\label{261}
\end{equation}

Here $dv_N^2(r) \equiv \partial_r [v_N^2(r)] \equiv [v_N^2(r)]'$
with $'$ denoting the differentiating with the argument $r$, or
$rx$.

Note that the part involving the integral with $dv_N^2(r/y)$ is
related to the information in the region $r' \ge r$. Here $r$ is
the point of the derived information such as $\mu(r)$, and $r'$ is
the source point of observation $v(r')$ in the integrand.
Therefore, this part with source function $r/y$ contains
information exterior to the target point $r$. On the contrary, the
source term with function of $rx$ represents the information
interior to the target point $r$.

Most of the time, the source information beyond certain
observation limit $r'=R$ becomes un-reliable or unavailable due to
the precision limit of the observation instruments. One will
therefore need to manually input the missing data in order to make
the integration result free of any singular contributions due to
the boundary effect. One will come back to this point in section
IV.

In addition, Eq. (\ref{261}) can be used to derive the total mass
distribution $M(r)$ of the spiral galaxy via the following
equation:
\begin{equation}
M(r)= 2 \pi \int_0^r r' dr' \; \mu(r').
\end{equation}

Note that the velocity function $v_N$ shown previously in this
paper is the rotation velocity needed to work with the total mass
of the system in the Newtonian dynamics. One can derive the
velocity $v(r)$ that works with the dynamics of MOND with the
relation given by Eq. (\ref{1}) and (\ref{2}). In short, $v \to
v_N$ in the limit of the Newtonian dynamics.

Throughout the rest of this paper, we will discuss the application
of these formulae both in the case of the Newtonian dynamics with
dark matter and in the case of MOND. Therefore, we will first
derive the velocity function $v_N(r)$ from $v(r)$ in the case of
MOND. As mentioned above, they follows the relation given by Eq.
(\ref{1}) and (\ref{2}).

Two different models will be studied later:

Case I: {Milgrom model}

Indeed, if one has $g(r)={v^2(r) / r}$ in the case of Milgrom
model, one can show that
\begin{equation} \label{32}
v_N^2(r)= { v^4(r) \over \sqrt{v^4(r) +g_0^2 r^2} }
\end{equation}
In dealing with the exterior part involving $r_0/y$, one has to
compute $dv_N^2(r)$ at large $r$. By assuming $v(r) \to v_R \equiv
v(r=R)$, one can show that:
\begin{equation}
dv_N^2(r \ge R) \to  - { v_R^4 g_0^2 r \over (v_R^4 +g_0^2
r^2)^{3/2}} .
\end{equation}
Here $R$ is the radius of the luminous galactic boundary. Mostly,
the flatten region of RC becomes manifest beyond $r \ge R$. In
addition, $v_N(rx)$ and $v_N(r/y)$ take the following form:
\begin{equation}\label{34}
v_N^2(rx)= { v^4 (rx) \over \sqrt{v^4 (rx) +g_0^2 r^2x^2} },
\end{equation}
\begin{equation} \label{35}
v_N^2({r \over y})= { v^4 ({r / y}) y  \over \sqrt{v^4({r / y})
y^2+ g_0^2 r^2 } }.
\end{equation}

Therefore, the surface mass density $\mu(r)$ can be written as,
with the velocity $v_N(r)$ given above,
\begin{eqnarray} \label{36}
&& \mu(r)={1 \over 2 \pi G}  \int_0^\infty
\partial_{r'} [{ v^4 \over \sqrt{v^4
+g_0^2 r'^2} }] \; H(r, r')dr'
\end{eqnarray}
in the case of Milgrom model.

Case II: {FB model}

Similarly, if one has $g(r)={v^2(r) / r}$ in the case of FB model,
one can show that
\begin{equation} \label{FB32}
v_N^2(r)= { v^4(r) \over v^2(r) +g_0 r }
\end{equation}
By assuming $v(r) \to v_R \equiv v(r=R)$, one can also show that:
\begin{equation}\label{FB33}
dv_N^2(r \ge R) \to  - { v_R^4 g_0 \over (v_R^2 +g_0 r)^2} .
\end{equation}
 In addition, $v_N(rx)$ and $v_N(r/y)$ take the following form:
\begin{equation}\label{FB34}
v_N^2(rx)= { v^4 (rx) \over {v^2 (rx) +g_0 rx} },
\end{equation}
\begin{equation} \label{FB35}
v_N^2({r \over y})= { v^4 ({r / y}) y  \over {v^2({r / y}) y+ g_0
r } }.
\end{equation}

Therefore, the surface mass density $\mu(r)$ can be written as,
with the velocity $v_N(r)$ given above,
\begin{eqnarray} \label{FB36}
&& \mu(r)={1 \over 2 \pi G}  \int_0^\infty
\partial_{r'} [{ v^4 \over v^2
+g_0 r' }] \; H(r, r')dr'
\end{eqnarray}
in the case of FB model. We will try to estimate the exterior
contribution of these two models shortly.

As mentioned above, it is easier to handle the numerical
evaluation involving regular function $E(x)$ instead of the
singular function $K(x)$. Therefore, one can perform an
integration-by-part and convert the integral in Eq. (\ref{18})
into an integral free of singular function $K(x)$. The result
reads, with $\mu(r)=\Theta(r)/( \pi^2 Gr)$,
\begin{equation}
\Theta(r)= \int_0^1 {V_N^2({r / x})-x V_N^2(r x) \over 1-x^2} E(x)
dx -\int_0^1 E'(x) V_N^2(rx) dx
\end{equation}
The last term on the right hand side of above equation can be
integrated by part again to give
\begin{eqnarray}
\Theta(r) &=& \int_0^1 {V_N^2({r / x})-x V_N^2(r x) \over
1-x^2} E(x) dx \nonumber \\
&+& \int_0^1 E(x) \partial_x V_N^2(rx) dx -V_N^2(r) .
\end{eqnarray}
Hence one has
\begin{eqnarray} \mu(r) &=& {1 \over \pi^2
Gr} [ \int_0^1 {V_N^2({r / x})-x V_N^2(r x) \over
1-x^2} E(x) dx \nonumber \\
&+& \int_0^1 E(x) \partial_x V_N^2(rx) dx -V_N^2(r) ] \label{39}.
\end{eqnarray}
Note again that the integral involving $v_N(rx)$ carries the
information $r' \le r$ while the integral with $v_N(r/x)$
represents the contribution from $r'> r$ by the fact that $0 \le x
\le 1$. As promised, one has transformed the singular $K$ function
into the regular $E$ function.

Although there are still singular contribution like $1/(1-x^2)$ in
the integrand, it is easier to handled since we know these
functions better than $K$ function. This is because we only know a
formal definition of this function via a set of definitions. Even
we have a rough picture about the form of $K(x)$ and $K'(x)$.
Numerical and analytical analysis could be difficult as compared
to dealing with the more well-known function like $1/(1-x^2)$. We
will show explicitly one of the advantage of this equation in next
section when one is trying to estimate the contribution from a
model describing the missing part of observation.

\section{contribution from the asymptotic region}

In practice, measurement in far out region is normally difficult
and unable to provide us with reliable information beyond the
sensitivity limit of the observation instrument. One often can
only measure energy flux and rotation curve within a few hundred
$kpc$ from the center of the galaxy. Beyond that scale of range,
signal is normally too weak to obtain any reliable data.
Therefore, one has to rely on various models to interpolate the
required information further out.

It it known that, contrary to the spherically system, exterior
mass does contribute to the inner region. Therefore, it is
important to estimate the exterior contribution carefully with
various models. In this section, we will study a velocity model
with a flat asymptotic form and its contribution to the inner part
in both Newtonian dynamics and MOND cases. One of the purpose of
doing this is to demonstrate the advantage of the regular function
formulae one derived earlier.

For a highly flatten galaxy, formulae obtained earlier in previous
section has been shown to be a very useful tool to predict the
dynamics of spiral galaxies. It is also a good tool for error
estimation. Possible deviation due to the interpolating data can
be estimated analytically more easily with the equations involving
only regular elliptic function $E(x)$. Note again that another
advantage of Eq. (\ref{39}) is that the $r$-dependence of the mass
density $\mu(r)$ has been extracted to the function $v_N$. This
will make the analytical analysis easier too.

Assuming that the observation data $v$ is only known for the
region $r \le R $, the following part of Eq. (\ref{39})
\begin{equation} \label{40}
\delta \mu(r \le R) ={1 \over \pi^2 G r}  \int_0^{r/R} dx  \left [
{v_N^2(r/x)  \over 1-x^2} \; E(x)  \right ]
\end{equation}
represents the contribution of $\mu(r\le R)$ from the unavailable
data $v_N(r')$ beyond the point $r = R$. To be more precise, one
can write $\mu(r \le R) = \mu_<(r)+ \delta \mu(r)$ with the mass
density $\mu_< (r)$ being contributed solely from the
$v_N(r')=v_N(r/x)$ data between $0 \le r' \le R$ or equivalently
$r/R \le x \le 1$. Explicitly, $\mu_<(r)$ can be expressed as
\begin{eqnarray} \mu_<(r) &=& {1 \over \pi^2
Gr} [   \int_{r/R}^1 {V_N^2({r / x}) \over
1-x^2} E(x) dx -\int_0^1 {x V_N^2(r x) \over 1-x^2} E(x) dx  \nonumber \\
&+& \int_0^1 E(x) \partial_x V_N^2(rx) dx -V_N^2(r) ] \label{411}.
\end{eqnarray}

As mentioned above, one will need a model for the unavailable data
to estimate the contribution shown in Eq. (\ref{40}). We will show
that a simple cutoff with $v(r
>R)=0$ will introduce a logarithmical divergence to the surface density $\mu(r=R)$.
The divergence is derived from the singular denominator $1-x^2$ in
Eq. (\ref{39}). The factor $1/(1-x^2)$ diverges at $x=1$ or
equivalently $r'=r$. A smooth velocity function $v_N(r)$
connecting the region $R-\epsilon < r < R +\epsilon$ is required
to make the combination $[V_N^2({r/x})-x V_N^2(r x)] / (1-x^2)$
regular at $x=1$.  Here $\epsilon$ is an infinitesimal constant.

Case I: Milgrom model

Let us study first the case of Milgrom model. Since most spiral
galaxies has a flat rotation curve $v(r \gg R) \to v_R $ with a
constant velocity $v_R$. For our purpose, let us assume that $v_R
= v(r=R)$ for simplicity. Hence the deviation (\ref{40}) can be
evaluated accordingly. Note again that this simple model agrees
very well with many known spirals.

 Note first that the Newtonian velocity $v_N(r)$ is given by
Eq.s (\ref{34}) and (\ref{35}) with $v(r >R)=V_R$. After some
algebra, one can show that
\begin{eqnarray} \label{451}
&& \delta \mu_M(r<R) = { 1 \over \pi^2 G r}  \int_0^{r/R} dx \left
[
{v_N^2(r/x)  \over 1-x^2} \;   \right ]E(x) \\
&=& {E_M \over \pi^2 G r}  \int_0^{r/R} dx \left [ {v_N^2(r/x)
\over 1-x^2} \;   \right ] \\  &=& {E_M v_R^4 \over \pi^2 Gr
\sqrt{v_R^4 +g_0^2r^2}}\nonumber \\ && \times \left [ \ln {r
\sqrt{v_R^4 +g_0^2R^2} + R \sqrt{v_R^4 +g_0^2r^2} \over
(g_0r+\sqrt{v_R^4 +g_0^2r^2})\sqrt{R^2-r^2}}
 \right ]\label{42}
\end{eqnarray}
Note that $\pi/2 \ge E(x) \ge 1$ is a monotonically decreasing
function with a rather smooth slope. The rest of the integrand is
also positive definite. Therefore, one can evaluate the integral
by applying the mean value theorem for the integral (\ref{451})
with $E_M \equiv E(x=x_M)$ the averaged value of $E(x)$ evaluated
at $x_M$ somewhere in the range $0 \le x_M \le 1$.

Case II: FB model.

Let us study instead the case of FB model. Let us also assume that
$v_R = v(r=R)$ for simplicity.  Note first that the Newtonian
velocity $v_N(r)$ is given instead by Eq.s (\ref{FB34}) and
(\ref{FB35}) with $v(r >R)=V_R$. After some algebra, one can show
that
\begin{eqnarray} \label{FB451}
&& \delta \mu_F(r<R) = {E_F \over \pi^2 G r}  \int_0^{r/R} dx
\left [ {v_N^2(r/x) \over 1-x^2} \;   \right ] \nonumber \\  &=&
{{E_F v_R^4 \over \pi^2 G r} } \int_0^{r/R} dx \left [ {1 \over
(1-x^2)(v_R^2 +g_0r/x)} \; \right ] \equiv  {{E_F v_R^2 \over
\pi^2 G r} } I \label{FB42}
\end{eqnarray}
with
\begin{equation}
I= \int_0^{r/R} dx \left [ {x \over (1-x^2)(x +g_0r/v_R^2)} \;
\right ]
\end{equation}
Note that $\pi/2 \ge E(x) \ge 1$ is a monotonically decreasing
function with a rather smooth slope. The rest of the integrand is
also positive definite. Therefore, one can evaluate the integral
by applying the mean value theorem for the integral (\ref{FB451})
with $E_F \equiv E(x=x_F)$ the averaged value of $E(x)$ evaluated
at $x_F$ somewhere in the range $0 \le x_F \le 1$. After some
algebra, one can show that
\begin{equation}
\delta \mu_F(r<R) = {{E_F v_R^4 \over 2 \pi^2 G r} } \left [ {1
\over v_R^2-g_0r} \ln (1+r/R) - {1 \over v_R^2+g_0r} \ln (1-r/R) -
{2g_0r \over v_R^4-g_0^2r^2} \ln [1+v_R^2/(g_0R)] \right ].
\end{equation}

Case III: Newtonian case.

Similarly, one can also evaluate the mass density in the case of
Newtonian dynamics with dark matter in need. Let us assume $v_N(r
\ge R) = v_R$ for simplicity again. As a result the deviation of
mass density required to generate the rotation curve $v_N(r) =
v(r)$ can be shown to be
\begin{eqnarray}
 \delta \mu_N(r<R) &=& {E_N \over \pi^2 G r}  \int_0^{r/R} dx
\left [ {v_N^2(r/x)  \over 1-x^2} \;   \right ] \nonumber \\
&=& {E_N v_R^2 \over 2 \pi^2 Gr } \ln {R+r \over R-r} \label{dmuN}
\end{eqnarray}
after some algebra. Note that, in deriving above equation, one
also applies the mean value theorem with $E_N \equiv E(x=x_N)$ the
averaged value of $E(x)$ evaluated at $x_N$ somewhere in the range
$0 \le x_N \le 1$.

To summarize, one has shown clearly with a simple model for the
asymptotic rotation velocity that formulae with regular $E$
function appears to be easier for analysis. This is mainly due to
the fact that $E(x)$ is smoothly and monotonically decreasing
function within the whole domain $x \in [0,1]$. In most cases,
mean value theorem is very useful in both numerical and analytical
evaluations.

{In addition, one notes that the mild logarithm divergent terms
appeared in the above final results are due to the cut-off at
$r=R$. A negative and equal contribution from the interior data
will cancel this singularity at $r=R$. To be more precisely, if we
turn off the $v$ function abruptly starting the point $r=R$ by
ignoring the exterior region contribution, a logarithmic
divergence will show up at $r=R$ accordingly. The appearance of
the logarithm divergence also emphasize that the boundary
condition of these physical observables at $r=R$ should be taken
care of carefully to avoid these unphysical divergences. In
practice, one normally adds a quickly decreasing $v_N(r >R)$ to
account for the missing pieces of information and to avoid this
singularity. Numerical computation may, however, bring up a small
peak near the boundary $r=R$ if the matching of $v_N$ at the
cutoff is not smooth enough. One will also show that similar
singular behavior also appears in the final expression of the
velocity function derived from a given data of mass distribution
in next section. Evidence also shows that exterior contribution
should be treated carefully in order to provide a meaningful
fitting result.}

In order to compare the difference of $\delta \mu$ for different
models, we find it is convenient to write $B \equiv v_R^2/(g_0R)$
and $s\equiv r/R$ such that $A$ and $r'$ both become dimensionless
parameters. Note that $B \sim 1.1$ if we take $v_R \sim 250\,
km/s$ and $R \sim 4.97 \times 10^4 \, ly$ from the data of Milky
Way. Therefore,  $B \sim 1.1$ is typically a number slightly
larger than 1. Hence, one can write above equations as
\begin{eqnarray}\label{A42}
\delta \mu_M(s<1) &=&  {E_M B^2 g_0 \over \pi^2 Gs \sqrt{B^2
+s^2}}\left [ \ln {s \sqrt{B^2+1} +  \sqrt{B^2 +s^2} \over
(s+\sqrt{ B^2+s^2})\sqrt{1-s^2}}
 \right ], \\
\delta \mu_F(s<1)  &=& {E_F B^2 g_0 \over 2 \pi^2 Gs }\left [ {
\ln (1+s) \over B-s} - {\ln (1-s) \over B+s}  + {2s \ln (1+B)
\over s^2-B^2}  \right ], \\
 \delta \mu_N(s<1)
&=& {E_N B g_0\over 2 \pi^2 Gs } \ln {1+s \over 1-s}.
\label{AdmuN}
\end{eqnarray}
In addition, one can estimate the deviation $\delta \mu$ at small
$s$ where $s \ll 1$, or $r \ll R$. The leading terms read:
\begin{eqnarray}\label{A79}
\delta \mu_M(s \ll 1) &=&  {E_M  g_0 \over \pi^2 G} \left [
\sqrt{B^2+1} -1  \right ] + {\rm O}(s), \\
\delta \mu_F(s \ll 1)  &=& {E_F  g_0 \over  \pi^2 G } \left [ B-
\ln (B+1) \right ] + {\rm O}(s), \\
 \delta \mu_N(s \ll 1 ) + {\rm O}(s).
&=& {E_N g_0\over  \pi^2 G }B   \label{A81}
\end{eqnarray}
at small $r$.  Note that $g_0/G \sim 0.18$. Therefore, the most
important contribution from the exterior contribution is near the
boundary at $r=R$. Our result shows that special care must be
taken near the boundary of available data. Appropriate matching
data beyond this boundary is needed to eliminated the naive
logarithm divergence. The compact expression also made reliable
estimation of the deviation possible.

\section{gravitational field derived from a given mass density}

One can measure the flux from a spiral galaxy and try to obtain
the mass density with the $M/L=$ constant law \citep{tf77}. Even
the $M/L$ law is more or less an empirical law, it does help us
with a fair estimate of the mass density distribution. We will
focus again on the physics of a highly flattened spiral galaxy.
Once the mass density function is known, for the range $ 0 \le r
\le R$, one can also compute the gravitational field $g_N(r)$ from
a given mass density function $\mu(r)$. Once the function $g_N(r)$
is derived, one can derive $g(r)$ following the relation given by
Eq.s (\ref{1}) and (\ref{2}) in the case of MOND.

Indeed, one can show that $g_N(r)$ is given by
\begin{equation}
g_N (r) =2 \pi G \int_0^\infty k dk \; \int_0^\infty r' dr' \;
\mu(r') J_0(kr')  J_1(kr)
\end{equation}
from the Eq.s (\ref{4}) and (\ref{9}).
 Therefore one has
\begin{equation}
g_N (r) =2 \pi G \int_0^\infty r' dr' \; \mu(r') H_1(r, r')
\end{equation}
with
\begin{equation}
H_1(r, r')=\int_0^\infty k dk J_1(kr) J_0(kr')=-\partial_r H(r,
r').
\end{equation}
By differentiating Eq. (\ref{161}), one can further show that
$H_1(r, r')$ becomes
\begin{eqnarray}
H_1(r,r') &=& {2 \over \pi r r'} \left [ K({r \over r'}) - {r'^2
\over r'^2-r^2} E({r \over r'}) \right ] \; { \rm for} \; r<r'
\label{47} \\
&=& {2 \over \pi (r^2-r'^2)} E({r' \over r}) \; \; { \rm for} \;
r> r' . \label{48}
\end{eqnarray}
Note that one has used the differential equations obeyed by $E$
and $K$ shown in Eq.s (\ref{24})-(\ref{26}). Following the method
shown in section IV, one can rewrite the equation as
\begin{eqnarray}
&& g_N(r)=4G \int_0^1  dx  \nonumber \\ && \left [ \mu(rx)  {x
E(x) \over 1-x^2} + \mu({r / x}) [{K(x) \over x^2} - {E(x) \over
x^2(1-x^2)}] \; \right ] \label{gNK}
\end{eqnarray}
after some algebra. In order to eliminate the singular function
$K(x)$, we can also convert $K(x)$ into regular function $E(x)$
following similar method. The result is
\begin{eqnarray}\label{50}
&& g_N(r)=4G \int_0^1  dx   \\ && \left [ \mu(rx)  {x E(x) \over
1-x^2} - \mu({r / x}) [{E(x) \over 1-x^2} + {E'(x) \over x}  ] \;
\right ] \nonumber.
\end{eqnarray}
With an integration-by-part, one can convert $E'(x)$ to a regular
function $E(x)$. The result is
\begin{eqnarray}\label{51}
&& g_N(r)=2G \mu_*(r) - 2 \pi G \mu (r)+ 4G \int_0^1  dx   \\ &&
\left [ {E(x) \over x^2(1-x^2)} [x^3 \mu(rx)- \mu({r / x})] -
{E(x) r \mu'(r/x) \over x^3}  \; \right ]  \nonumber.
\end{eqnarray}
Here $\mu_*(r) \equiv \lim_{x \to 0} 2 \mu(r/x)/x$. If $\mu(r \to
\infty)$ goes to 0 faster than the divergent rate of $r$,
$\mu_*(r)$ will vanish or remain finite. For example, if
$\mu(r>R)= 2\mu_R Rr /( R^2+r^2)$, one can show that $\mu_*(r) = 4
\mu_RR/r$. Here $\mu_R \equiv \mu(r=R)$. We will be back with this
model in a moment.

Similar to the argument shown in section IV, one can show that the
terms with $\mu(r/x)$ in Eq. (\ref{51}) will contribute
\begin{eqnarray}\label{52}
\delta g_{N0}(r\le R )=-4G \int_0^{r/R}  dx  E(x) \left [ {
\mu(r/x ) \over x^2 (1-x^2)}  + {r \mu'(r/x) \over x^3} \right ]
\end{eqnarray}
to the function $g_N(r)$ due to the unavailable data $\mu(r > R)$.
Note, however, that part of the exterior region contribution has
been evaluated via the integration-by-part process in deriving Eq.
(\ref{51}) involving $v_N(r/x)$ and $E'(x)$. Therefore, one should
start with the complete version (\ref{50}) in evaluating the
deviation of $g_N$ due to the exterior part. Hence one should have
\begin{eqnarray}\label{53}
\delta g_N(r \le R)=-4G \int_0^{r/R}  dx  \mu({r \over x})  \left
[ {E(x) \over 1-x^2} + {E'(x) \over x} \right ] .
\end{eqnarray}
For simplicity, we will assume the following form of mass
distribution in the region $r >R$,
\begin{equation} \mu(r>R)= \mu_R {2 Rr \over
R^2 +r^2}
\end{equation}
with $\mu_R \equiv \mu(r=R)$. Note that continuity of $\mu(r=R)$
across the matching point $r=R$ is managed to remain valid in this
model. Moreover, $\mu(r \to \infty) \to 0$ is expected to hold for
the luminous mass of the spiral galaxies. After some algebra, one
can show that
\begin{eqnarray}\label{55}
 \delta g_N(r)= -4G \mu_R {R \over r}\left [ E({r \over R})-\pi +
A(r) \right ]
\end{eqnarray}
with $A(r)$ given by the integral
\begin{eqnarray} \label{61}
A &=& \int_0^{1/b} dy \left [ {1 \over (1-y)(1+by)} + {2b \over
(1+by)^2} \right ] E(\sqrt{y})  \\
&=& E_1 \int_0^{1/b} dy \left [ {1 \over (1-y)(1+by)} + {2b \over
(1+by)^2} \right ]
\end{eqnarray}
with the first two terms in Eq. (\ref{55}) the contribution from
integration-by-part of $E'(x)$. Here one has defined $y=x^2$ and
$b = R^2/r^2$ for convenience.

In addition, one also applies the mean value theorem to the
integral (\ref{61}) by noting again that (i) $\pi/2 \ge E(x) \ge
1$ is a monotonically decreasing function with a rather smooth
slope, and (ii) the integrand is a positive function throughout
the integration range. Therefore, the integral $A(r)$ can be
evaluated by applying the mean value theorem with $E_1 \equiv
E(x=x_1)$ the averaged value of $E(x)$ evaluated at $x_1$
somewhere in the range $0 \le x_1 \le 1$.

The remaining integral in $A$ can be evaluated in a
straightforward way and one finally has
\begin{eqnarray}\label{dgN}
&& \delta g_N(r)= \\ && -4G \mu_R {R \over r}\left [ E_1 [1+ {r^2
\over R^2 +r^2} \ln {2 R^2 \over R^2 - r^2 } ] +E({r \over R})-\pi
\right ] \nonumber .
\end{eqnarray}
This is the contribution to the Newtonian field $g_N(r \le R)$ due
to the unknown exterior mass contribution. Note that similar
singularity also appears at $\delta g_N(r=R)$ which means that a
careful treatment in modelling the unknown exterior mass is
needed.

Case I: Milgrom model.

For the case of Milgrom model, one can show that
\begin{equation}
g(r) = \sqrt{{g_N^2(r) + \sqrt{g_N^4(r) +4 g_N^2(r)g_0^2} \over 2
} }.
\end{equation}
Therefore, the deviation $\delta g^2(r)$ can be shown to be
\begin{equation}
\delta g^2(r) = \left [ { 1 \over 2} + {g_N^2(r) + 2 g_0^2 \over 2
\sqrt{g_N^4(r) +4 g_N^2(r)g_0^2}  } \right ] \delta g_N^2(r)
\end{equation}
by solving the algebraic equations (\ref{1})-(\ref{2}).

Case II: FB model.

For the FB model, one can show that
\begin{equation} \label{FB04}
g={ \sqrt{{g_N}^2+4g_0g_N} +g_N \over 2} .
\end{equation}
Therefore, the deviation $\delta g^2(r)$ can be shown to be
\begin{equation}
\delta g(r) = { 1 \over 2} \left [ 1 + {g_N(r) + 2 g_0 \over
\sqrt{g_N^2(r) +4 g_N(r)g_0}  } \right ] \delta g_N(r)
\end{equation}
by solving the algebraic equations (\ref{FB04}).

Therefore, the deviation $\delta g(r)$ can be evaluated from above
equation to the first order in $\delta g_N(r)$ with all $g_N(r)$
replaced by $g_{N<}(r)$. Here $g_{N<}(r)$ is defined as the
contribution of interior mass to the Newtonian field $g_N(r)$,
namely,
\begin{equation}
g_N(r \le R)=g_{N<}(r) + \delta g_N(r) .
\end{equation}

To summarize again, one has shown that the formulae with regular
$E$ function appears to be helpful in deriving the gravitational
field strength for numerical and analytical purpose.

\section{conclusion}

We have reviewed briefly how to obtain the surface mass density
$\mu(r)$ from a given Newtonian gravitational field $g_N$ with the
help of the elliptic function $K(r)$ in this paper. The integral
involving the Bessel functions is derived in detailed for
heuristical reasons in this paper too. In addition, a series of
integrable model in the case of MOND is also presented in this
paper.

One has also converted these formulae into a simpler compact
integral making numerical integration more accessible and
analytical estimate possible in this paper. The apparently
singular elliptic function $K(r)$ is also converted to
combinations of regular elliptic function $E(r)$ by properly
managed integration-by-part.

As a physical application, one derives the interior mass
contribution $\mu(r<R)$ from the possibly unreliable data $v(r
>R)$ both in the cases of MOND and in the Newtonian dynamics.
Detailed results are presented for Milgrom model and FB model for
the case of MOND. In particular, the singularity embedded in these
formulae are shown to be a delicate problem requiring great
precaution. Similarly, one also tries to derive similar results
for the corresponding $g_N$ from a given $\mu(r)$. One also
presents a simple model of exterior mass density $\mu(r
>R)$ as a simple demonstration. The corresponding result in the theory of
MOND is also presented in this paper.

In section III, one has studied many solvable models in details.
Analysis is generalized to Newtonian models, Milgrom models as
well as the FB models. In practice, may convert the RC data from
$v$ to $v_N$ following the transformation formula of either
Milgrom model or FB model. For the case of Newtonian model, the RC
data gives exactly the Newtonian velocity, namely, $v=v_N$. One
can then fit the resulting $v_N^2$ in expansion of these modes in
order to analyze the RC in basis of of these basis modes. This
helps analytical understanding of the spiral galaxies more
transparently. The properties of each modes can be easily
understood because they are integrable. The corresponding
coefficients $C_{ni}$ and $a_j$ will determine the contributions
of each modes to any galaxies. One will then be able to construct
tables for spiral galaxies with the corresponding coefficients of
each modes. Hopefully, this expansion method will provide us a new
way to look at the major dynamics of the spiral galaxies.
Analytic approach to the dynamics of highly flattened galaxies

In summary, the compact expressions (\ref{40}) and (\ref{51}) have
been shown in this paper to be useful in the estimate of the mass
density $\mu(r <R)$ and $g(r <R)$ from the exterior data at $r'
>R$. Explicit models are presented in this paper.

One also presents a more detailed derivation involving the
definition of the elliptic functions $E$ and $K$. Various useful
formulae are also presented for heuristic purpose.

One also focuses on the application in the case of modified
Newtonian dynamics for two different successful models: the
Milgrom model and the FB model. The theory of MOND appears to be a
very successful model representing possible alternative to the
dark matter approach. Nonetheless, MOND could also be the
collective effect of some quantum fields under active
investigations \citep{mond1, mond2}. The method and examples shown
in this paper should be of help in resolving the quested puzzle.

\acknowledgments
 {\bf \large Acknowledgments} This work is supported in
part by the National Science Council of Taiwan. We thank professor
Zhao H.S. for helpful suggestions and comments.


\end{document}